\documentclass{aa}
\usepackage{graphicx}
\input{epsf}

\def\be{\begin{equation}}
\def\ee{\end{equation}}
\def\ba{\begin{eqnarray}}
\def\ea{\end{eqnarray}}

\def\msun{M_\odot}

\def\ltsima{$\; \buildrel < \over \sim \;$}
\def\simlt{\lower.5ex\hbox{\ltsima}}
\def\gtsima{$\; \buildrel > \over \sim \;$}
\def\simgt{\lower.5ex\hbox{\gtsima}}

\begin{document}

\title{Primordial star formation triggered by UV photons from UHECR}

\author{Yu. A. Shchekinov\inst{} and E. O. Vasiliev\inst{}}
\offprints{Yu. A. Shchekinov}

\institute{Department of Physics, University of Rostov, %\\
and Isaac Newton Institute of Chile, Rostov-on-Don Branch,\\
Rostov-on-Don, 344090 Russia\\
\email{yus@phys.rsu.ru, eugstar@mail.ru}
}
\date{Received 25 October 2003 / Accepted 6 January 2004}

\titlerunning{Primordial star formation}

\abstract{The presence of ultra-high energy cosmic rays in the universe in the pre-ionization epoch enhances significantly H$_2$ kinetics in virialized halos. It results in more efficient radiative cooling and decreases the lower mass limit of the first star-forming systems. Consequently, the fraction of baryons contained in the first luminous objects and their contribution to the reionization of the universe can significantly increase in comparison with the standard scenario.
\keywords{Cosmology: early universe, cosmology: miscellaneous, stars: formation}}

\maketitle

\section{Introduction}
%\label{1}
In hierarchical cold dark matter (CDM) scenarios the formation of the first stellar 
objects involves dark matter halos of total (dark and baryonic matter) masses $M_h\sim 2\times 10^6\msun$ at redshifts $z\sim 20$ (Tegmark et. al. 1997, hereafter T97), or lower ($M_h\sim 2\times 10^5\msun$) if the effects of violent relaxation of dark halos are taken into account 
(Vasiliev \& Shchekinov 2003). This process, and the corresponding halo 
masses where the first stellar objects may form, is determined by the cooling 
of baryons mainly through radiation of thermal energy in rotational lines 
of H$_2$ molecules. In primordial gas, H$_2$ molecules can form only in 
reactions involving H$^-$ and H$_2^{+}$ ions, and therefore the H$_2$ yielded strongly depends on fractional ionization $x=n_e/n$ (Galli \& Palla 1998). 
In the standard cosmological recombination, H$_2$ can reach not more than 
$f_2=n({\rm H_2})/n=10^{-3}$ in already virialized halos and it determines a rather high lower mass limit $M_h$ (T97). However, in cosmologies with decaying dark matter particles, the recombination dynamics itself changes, as was first suggested by Sciama (1982, 1990), 
Adams, Sarkar \& Sciama (1998). As a result, H$_2$ chemistry can be 
significantly altered. 

Ultra-high energy cosmic rays (UHECR), if they formed in the so-called 
top-to-bottom scenario from decaying superheavy dark matter (SHDM) 
particles with masses $M_X\simgt 10^{12}$ GeV (Berezinsky, 
Kachelrie\ss \ \& Vilenkin, 1997, Kuzmin \& Rubakov, 1998, Birkel \& 
Sarkar, 1998), may also have a strong impact on cosmological recombination. 
A detailed description of the ionization history of the primeval plasma 
in the presence of such UHECR has been reported recently by Doroshkevich \& 
Naselsky (2002, hereafter DN), Doroshkevich et al. (2003). These authors conclude that the UHECR converted through electromangetic cascades into UV photons increase the 
hydrogen ionization at $z\sim 10-50$ by a factor of 5 to 10. Correspondingly, the 
fraction of molecular hydrogen in virializing halos can increase by factor 
of 10-20 depending on details of their dynamics. In this paper we discuss briefly
the effects of such enhanced kinetics of H$_2$ on minimal masses and evolution of early luminous objects. 

Throughout this paper we assume a $\Lambda$CDM cosmology with parameters 
$(\Omega_0, \Omega_{\Lambda}, \Omega_m, \Omega_b, h ) = (1.0,\ 0.71,\ 0.29,\ 0.047,\ 0.72 )$ as inferred from the Wilkinson Microwave Anisotropy Probe (WMAP), and a deuterium abundance of $2.62\times 10^{-5}$ (Spergel et al. 2003). 
 
In Section 2 we estimate minimal masses of early luminous objects. 
In Section 3 we analyze qualitatively how additional ionizing photons change thermal 
dynamics of individual contracting baryon condensations. In Section 4 
we discuss possible cosmological consequences. Section 5 summarizes our conclusions.

\section{Minimum masses of early luminous objects}

Formation of early luminous objects (stars, quasars or stellar systems) in the 
universe and their masses are determined by the ability of primordial gas 
to cool radiatively and to form self-gravitating contracting agglomerations 
of baryons. In the hierarchical CDM scenario, contraction of baryons is catalyzed by 
growing fluctuations of gravitational potential from virialized dark 
matter halos. The exact description requires a full 3D multi-fluid 
numerical approach, and although such a description explains the whole 
variety of dynamical processes involved, it remains still highly time 
consuming (Abel, Bryan \& Norman 2000, Bromm, Coppi \& Larson 2001, Abel, 
Bryan \& Norman 2002). For our purpose, to illustrate possible consequenses 
of the presence of extra ionizing photons generated by UHECR 
on the process of the initial star formation, one can follow a simplified 
approach suggested by T97. This approach is 
based on the solution of an ordinary differential equation for thermal energy 
of baryons confined by the gravitational potential of a virialized halo. An 
essential assumption of T97 is in their criterion for baryons to cool: 
they assumed that the baryons in a halo with mass $M$ will cool progressively,  
if the baryon temperature decreases sufficiently: $T(\eta z) \le \eta T(z)$ 
faster than in one comoving Hubble time $t_H(z)$, where $\eta = 0.75$. 

Vasiliev \& Shchekinov (2003) have used a different criterion which 
requires stronger conditions for baryons to cool and form a self-gravitating 
configuration: they assumed that it occurs if in the process of baryon 
contraction in the gravitational field of a dark halo, the baryons reach a mean density 
equal to the mean density of dark matter within the sphere occupied by baryons.  
Obviously, it results in an increase of the minimal baryon mass able to cool. 
However, to qualitatively understand possible effects from UHECR 
on star formation, here we present calculations of the minimal mass only  
within the T97 criterion. 

Consider first a single virialized halo of mass $M_h=10^6\msun$. 
The equations describing ionization fraction 
$x=n({\rm H}^+)/n$, molecular fraction $f=n({\rm H}_2)/n$ and thermal 
evolution of baryons read as (T97)
\begin{equation}
\label{el}
\dot x = -k_1 n x^2,
\end{equation}
\begin{equation}
\label{h2}
\dot f = k_m n(1-x-2f)x,
\end{equation}
\begin{equation}
\label{tem}
\dot T = {2\over 3}{\dot n \over n}T + \Lambda_{\rm C} - \Lambda_{\rm HI} - 
\Lambda_{\rm H_2},
\end{equation}
where $\Lambda_{\rm C},\ \Lambda_{\rm HI},\  \Lambda_{\rm H_2}$, 
$\Lambda_{\rm HD}$ are the Compton, HI, H$_2$ and HD cooling rates, respectively, $k_1$ and $k_m$ are the reduced reaction rates (see for details T97). A reduced HD chemistry has been also added to account for effects from HD cooling at temperatures lower than 400 K. We have run the system for a grid of halo masses $M_h$ with initial conditions corresponding 
to the fractions of species and the temperature at the turnaround redshift;
the evolution before the turnaround point has been found from the T97 model.

The rate of production of ionizing photons by UHECR can be written as
$dn_{\gamma}/dt = \epsilon(z)H(z)n(z)$
where $n_{\gamma}$ is the number density of ionizing photons, $H(z)$ is the
Hubble parameter and $n(z)$ the gas number density (see DN). Here the
efficiency $\epsilon(z)$ is defined as (DN):
\begin{equation}
\label{param}
\epsilon \simeq {2.5 \times 10^{-4} \over 1+z} {M_{16}}^{2-\alpha} \Theta_{tot},
\end{equation}
where $\alpha$ is the power index of the spectrum of injected photons for a decay of a single SHDM particle, $M_{16} = M_X/10^{16}~$GeV, $M_X$ is the mass of the SHDM particles. For the measurable distortions of the power spectrum of the CMB anisotropy $\Theta_{tot} \simeq 10^4{M_{16}}^{-0.5}$ (DN). In the calculations we assume for the efficiency $0.3/(1+z)$ as a  typical value. 

\begin{figure}
  \resizebox{\hsize}{!}{\includegraphics{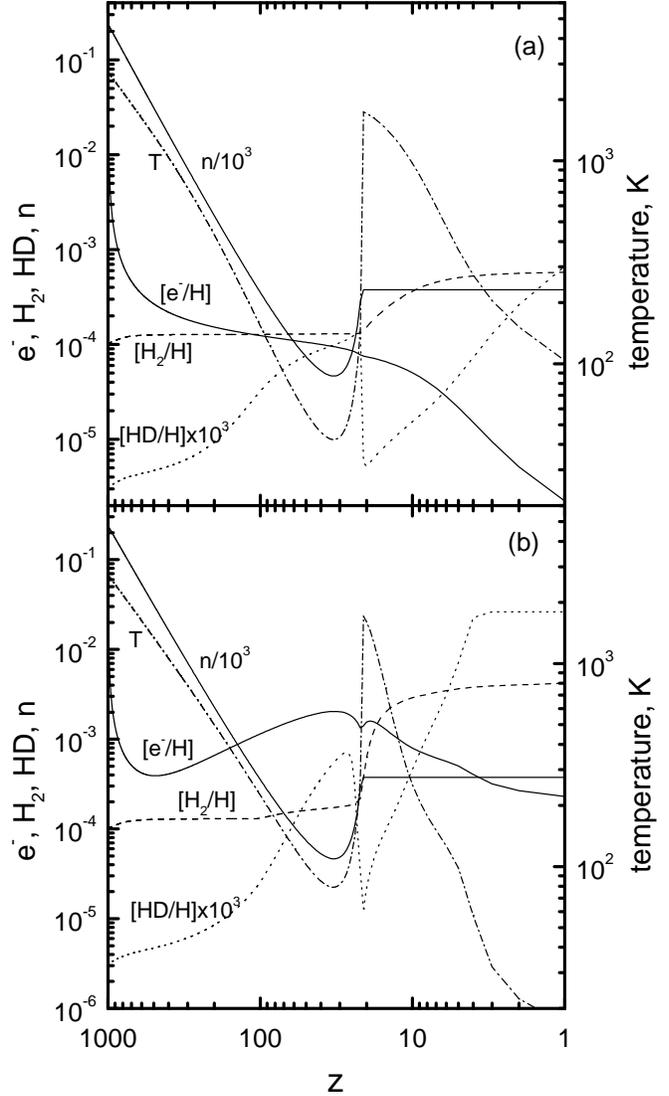}}
      \caption{The ionization (solid), molecular (dash) and HD (dot) fractions, 
               and the thermal evolution (dot-dash) of baryons in a virialized 
               halo with 
               mass $M_h=10^6\msun$. The upper panel depicts standard case 
               without additional Ly$c$ photons, while the lower panel represents 
               a model with extra UV photons with parameters corresponding to the 
               model $\epsilon=0.3/(1+z)$ in Doroshkevich \& Naselsky (2002). 
               The HD fraction is multiplied by $10^3$. 
               The number density $n$ is divided by $10^3$. }
         \label{Fig1}
\end{figure}

In Fig.1 the thermochemical evolution of baryons in a virialized
halo with mass $M_h=10^6\msun$ for the standard recombination scenario 
(upper panel) and that with an enhanced fractional ionization (lower panel) is shown. 
The small additional electron fraction produced by UHECR photons results in more efficient 
H$_2$ molecule formation, such that inside a collapsing low-mass halo the molecular hydrogen 
fraction reaches a value of about $(0.6-4)\times 10^{-3}$ depending on $\epsilon$. 
As a concequence the temperature drops faster, and this enhances formation of HD molecules, 
and practically all deuterium is bound into HD molecules when the temperature falls below 200 K. Thus, even for a relatively low production rate of ionizing photons 
$\epsilon$ the molecular cooling rate increases significantly. 

In principle, additional UV photons from UHECR can heat baryons in the halo, 
however in most cases their impact on thermal evolution inside a formed halo seems to 
be unimportant: the corresponding photoionization heating is $H_{UV}\sim 2\times10^{-29} \epsilon~n(1+z)^{1.5}$~erg~cm$^{-3}$~s$^{-1}$ ($n$ is the baryon density) and can be cooled by molecular hydrogen with $n($H$_2)/n \sim 3\times 10^{-4}$ at $T\sim 700$~K, i.e. smaller than virial temperature for halos with $M_h \geq 3\times 10^5\msun$. This means that when a halo of $M_h > 3\times 10^5\msun$ reaches its virial state with $T_v \geq 700$~K, radiative energy losses by H$_2$ molecules exceed UV heating. Note also that the rate destruction of H$^-$ ions by extra photons is only about $10^{-19}$s$^{-1}$, which is much smaller than the rate of convertion to H$_2$ $\sim 10^{-12}$s$^{-1}$.

In such conditions self-gravitating baryon configurations form more easily and have 
the lower masses. Fig. 2 shows the $z$-dependence of lower mass limit $M_{hl}(z)$ of 
halos able to form cooling self-gravitating baryon condensations. 
The thick solid line depicts the masses corresponding to 3$\sigma$ peaks in a $\Lambda$CDM model. Only masses in the interval $M_{hl}(z)<M_h<M_{3\sigma}(z)$ are supposed to form early luminous objects. In general, additional ionizing photons from UHECR can stimulate formation of early objects in the universe with masses of an order of magnitude lower than in conditions with the standard cosmological recombination.

\begin{figure}
  \resizebox{\hsize}{!}{\includegraphics{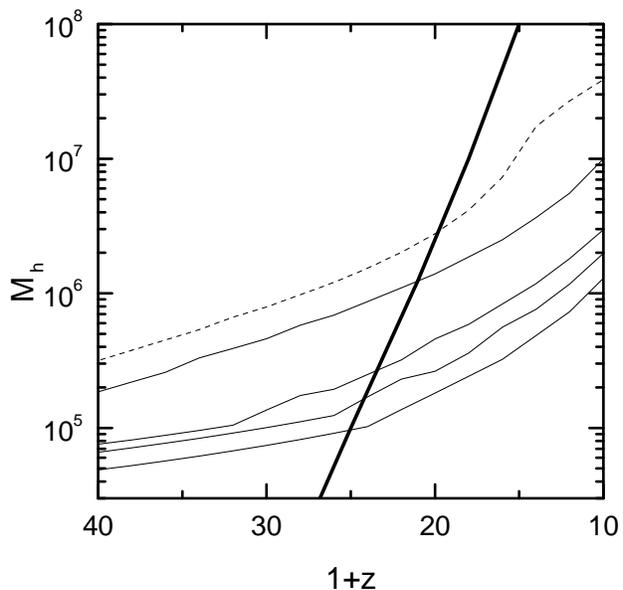}}
      \caption{Minimum mass able to cool and collapse. The dashed line depicts 
the lower mass limit estimates by T97 in standard cosmological recombination. The four thin 
solid lines represent the lower mass for the rate of ionizing photon production by 
UHECR, $\epsilon = (0.1, 0.3, 1, 3)/(1+z)$ (from top to bottom). The thick 
solid line corresponds to 3$\sigma$ peaks in a $\Lambda$CDM model.}
         \label{Fig2}
\end{figure}

\section{Evolution of individual halos}

Besides determining the minimal mass of the first baryon objects, the question 
of how individual halos of large masses evolve is also of interest for 
a qualitative understanding of the effects of UHECR on primordial star 
formation. Of course, the full description can be reached only in 
multi-dimensional numerical simulations, however, to understand qualitatively what one can expect from the presence of UHECR in the early universe for primeval star formation, a simplified consideration may be helpful. 

Within this approach we will describe baryons in a virialized halo by the 
following set of equations widely used in simplified calculations (see e.g., 
Tegmark, Silk \& Evrard 1993) 

\begin{equation}
\label{radius}
\dot R= u,
\end{equation}

\begin{equation}
\label{velocity}
\dot u={4kT\over \mu m_p R}-{4\over 3}\pi G(\rho_d+\rho)R,
\end{equation}

\begin{equation}
\label{temperature}
\dot T=-
{2kT\over \mu R}u
-\sum_i \Lambda_i,
\end{equation}
where all notations have their usual senses, $\Lambda_i$ are as in (\ref{tem}). 
These equations were solved with a complementary set of destruction for ionization 
and chemical kinetics accounting for additional ionization of H and 
photodestruction of ${\rm H^-}$ ions by UHECR photons, and a simplified 
treatement of effects of optical thickness on the corresponding cooling function by introducing an exponential factor $\exp(-\tau_{{\rm H}_2})$, where $\tau_{{\rm H}_2}$ is 
the average optical depth in ${\rm H}_2$ lines. We calculated the evolution of 
baryons for a halo with a given mass from its virialized 
state to the asymptotic state when concentrations of H$_2$ and HD reach their 
limiting values, and temperature settles onto a regime determined by adiabatic 
heating, radiative cooling and energy exchange with cosmic background radiation. 

The results are shown in Fig. 3, where temperature, baryon density
and Jeans mass within a baryon condensation are depicted, respectively for 
three halo masses: $M_h=10^6,~10^7$ and $10^8\msun$.  

\begin{figure}
  \resizebox{\hsize}{!}{\includegraphics{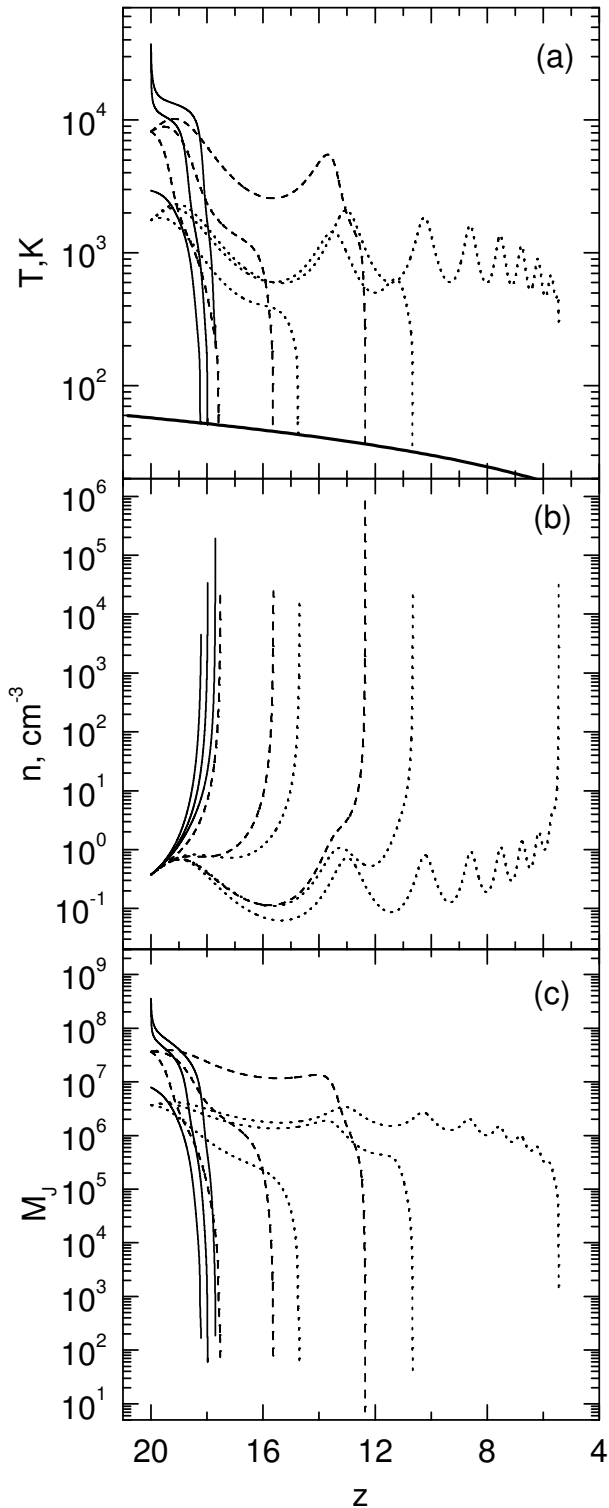}}
      \caption{Temperature (upper panel), density (middle panel) and 
       Jeans mass (lower panel) evolution of baryons in halos with 
       masses $10^6,~10^7$ and $10^8\msun$ virialized at $z=20$ for the standard 
       scenario, $\epsilon = 0$, and for cosmological recombination dominaned by 
       UHECR photons, $\epsilon = 0.1,~3$: {\it solid} lines correspond to halo mass 
       $M_h=10^8\msun$, {\it dashed} lines -- $M_h=10^7\msun$, 
       {\it dotted} lines -- $M_h=10^6\msun$. On panel (a) the temperature 
       of CMB is shown by the thick solid line.
              }
         \label{Fig3}
\end{figure}
The difference between the temperatures is readily seen: in the case 
$\epsilon=0$ (standard recombination), after an initial period of  
cooling, temperature starts growing for the halo mass $M_h=10^7\msun$ 
because strong adiabatic heating cannot be balanced by H$_2$ with 
H$_2$/H$\sim 3\times 10^{-4}$, and only in subsequent contraction when 
H$_2$/H increases fast and reaches $\simgt 5\times 10^{-4}$ temperature 
decreases again. It happens only at $z\simeq 13.5$. 
In the case of nonzero input from UHECR, due to enhanced production H$_2$ 
abundance reaches sufficiently high values $\sim 10^{-3}$ already at the initial 
stages, so that a powerful H$_2$-cooling sets 
in at the very initial contraction phase, and the temperature decreases 
monotonously down to $\sim 50$ K (the CMB temperature) 
in a much shorter time -- the corresponding redshift $z\simlt 16.5$ for 
$\epsilon>0.3$. At the latest phases of the contraction ($T\simlt 200$ K), 
the thermal regime is controlled by cooling from HD molecules whose 
abundance reaches the abundance of deuterium in models with $\epsilon
\simgt 0.3$. Halos with masses $M_h=10^8\msun$ show a similar behavior 
depending on $\epsilon$ (as shown by solid lines in Figs. 3), although
the difference between the models with $\epsilon=0$ and $\epsilon>0$ 
is less pronounced here. Thus, the effects of additional ionization 
are more important for the halos at the low end of masses. 

Generally, the Jeans mass within the baryonic core decreases faster 
when the ionization efficiency from UHECR photons increases. At later stages it reaches asymptotically the value $M_J\sim 100(1+z)^{3/2}~n^{-1/2}\msun$ while baryonic cores or fragments remain optically thin for HD lines, here the dependence on $z$ 
stems from asymptotical value of baryon temperature $T_b\simeq T_0(1+z)$. 
When the Jeans mass becomes opaque in HD lines, its value is 1-2~$\msun$. For the standard scenario ($\epsilon=0$), the Jeans mass at later stages of contraction 
is nealy coincident (within factor of 2) with the mass of a collapsing core $M_c\simeq 100~\msun$ reached in numerical simulations by Abel, Bryan \& Norman (2003). Note that in their simulation the temperature of gas does not decline to the CMB value -- the typical temperature in the core is $\sim 300$~K. Our results differ from the results of a full 3D simulation apparently due to contribution from HD molecules.

\section{Discussion}
An obvious consequence of the enhaced H$_2$ production discussed here is closely 
related to the post-dark age evolution of the universe: its reionization and 
enrichment with metals by the primary stellar nucleosynthesis. The presence of an additional ionization from UHECR photons decreases the lower limit for the first baryon condensations and shifts their formation to earlier epochs: with simple T97 
criterion for formation of baryon objects, additional UV photons decrease the 
minimal mass from $\sim 10^6~\msun$ to $\sim 10^5~\msun$, and correspondingly 
increase the redshift from $z\simeq 20$ to $z\simeq 25$. In a hierarchical scenario (e.g. Lacey \& Cole, 1993) the number of $10^6 \msun$ halos at redshift $z \approx 20$ is an order of magnitude smaller than of $10^5 \msun$. This means that the production of Ly-$c$ 
photons by the first stars may have been significantly greater. If the efficiency of star formation in smaller halos is equal to that in more massive halos, the production of photons increases by a factor of 2, which would be sufficient to reionize the universe at levels established by WMAP (see discussion in Cen 2003). However, if the efficiency of star formation depends on halo mass as $f_* \sim M^{\alpha}$, than with $\alpha = 0.8$ as suggested for instance by Kasuya et al (2003) the amount of Ly-$c$ photons increases only by $\sim 20~\%$. 

The more numerous first baryon objects expected in a model with the presence 
of UHECR photons may change significantly also the metal enrichment history 
of the intergalactic medium (IGM), 
as observed in Ly-$\alpha$ forest clouds. One of the key problems, related to 
the initial pollution of the IGM by metals is connected not only 
with their total amount observed in Ly-$\alpha$ absorbers, but with their 
widespread presence at high redshifts, and particularly in underdense 
regions far from possible galactic sources (see, the recent discussion in 
Schaye et al 2003). A possible solution of this problem can be related to the  
highly overlapping metal -- enriched outflows produced by the early star-forming 
halos, as suggested by Nath \& Trentham (1997), Madau et al (2001), 
Scannapieco et al (2002). In particular Scannapieco et al (2002) predict that 
halos of intermediate masses of $3\times 10^{8-9}~\msun$ with a star 
formation rate corresponding to a starburst mode can explain basic features 
of the metal distribution at high $z$. By accounting for the fact that in the
presence of UHECR the number density of lower massive star-forming 
halos can be an order of magnitude higher, and star formation 
in them sets in earlier, even much lower (factor of 20) star formation activity than that 
assumed in Scannapieco et al. (2002) can match the observed evolution of metals 
in the IGM. 

\section{Conclusions}
%\begin{spacing}{2}
We have considered the impact of additional ionizing photons from UHECR on the lower mass limit of the early luminous objects. We showed that in the presence of UHECR the early luminous objects are an order of magnitude less massive, and form earlier in comparison to the predictions of the standard recombination history. In particular, for $\epsilon=0.3$ the objects with baryonic mass $3\times 10^5~\Omega_b~\msun$ can form at redshift $z=25$. Thus, under these conditions the baryons become able to cool and form luminous objects at higher redshifts and their fraction increases by an order of magnitude. A possible consequence can be a larger and earlier production of ionizing photons in the primary stellar nucleosynthesis required for the reionization of the universe.
%\end{spacing}

\section{Acknowledgements}
We thank the referee D. Galli for valuable suggestions.

\end{document}